\documentclass[a4paper,prc,unsortedaddress,superscriptaddress,twocolumn,floatfix,nofootinbib]{revtex4}
\usepackage{pstricks}
\usepackage{graphicx}
\usepackage{bm}
\usepackage{color}
\usepackage[normalem]{ulem}%Added by Martin, to cross out words with the command \sout{}

\newcommand{\ipnl}{Institut de Physique Nucl\'eaire de Lyon,
69622 Villeurbanne, France}
\newcommand{\nscl}{National Superconducting Cyclotron Laboratory and
Department of Physics and Astronomy,
Michigan State University, East Lansing 48824 MI, USA}
\begin{document}
\title{Kinematic sensitivity to the Fierz term of $\beta$-decay differential spectra} 

%%% quote always \altaffiliaton before \affiliation

\author{M.~Gonz\'alez-Alonso}
\altaffiliation[E-mail: ]{mgonzalez@ipnl.in2p3.fr}
\affiliation{\ipnl}

\author{O.~Naviliat-Cuncic}
\altaffiliation[E-mail: ]{naviliat@nscl.msu.edu}
\affiliation{\nscl}

%%%%%%%%%%%%%%%%%%%%%%%%%%%%%%%%%%%%%%%%%%%%%%%%%%%%%%%%%%%%%%%%%%%%%%%%%%%%%%%%
\date{\today}
%\date{Set fixed date here}
%%%%%%%%%%%%%%%%%%%%%%%%%%%%%%%%%%%%%%%%%%%%%%%%%%%%%%%%%%%%%%%%%%%%%%%%%%%%%%%%
\begin{abstract}
The current most stringent constraints on exotic scalar or tensor couplings
in neutron and nuclear $\beta$ decay, involving left-handed neutrinos,
are obtained from the Fierz interference term.
The sensitivity to this term in a correlation coefficient is usually driven
by an energy-averaged kinematic factor that increases monotonically toward
smaller values of the
$\beta$ endpoint energies. We first point out here that this property does
not hold for certain differential observables that are directly sensitive to
the Fierz term, such as the $\beta$ or the recoil energy spectrum. 
This observation is relevant for the selection of sensitive transitions in
searches for exotic couplings through spectrum shape measurements.
We then point out previous errors in the exploitation of measurements of
the $\beta-\nu$ angular correlation coefficient and discuss their impact on the extraction of constraints on exotic couplings.
\end{abstract}

%\pacs{}% AIP is changing from PACS codes

%%%%%%%%%%%%%%%%%%%%%%%%%%%%%%%%%%%%%%%%%%%%%%%%%%%%%%%%%%%%%%%%%%%%%%%%%%%%%%%%
\maketitle
%%%%%%%%%%%%%%%%%%%%%%%%%%%%%%%%%%%%%%%%%%%%%%%%%%%%%%%%%%%%%%%%%%%%%%%%%%%%%%%%
\section{Introduction}

Precision measurements in nuclear and neutron decays have played a
crucial role in the development of the
$V-A$ theory of the weak
interaction, which is embedded in the framework of the standard
electroweak model (SM).
Today, these experiments constitute sensitive probes to ``exotic'' currents,
such as right-handed vector currents or scalar and tensor currents, resulting from the exchange of new heavy bosons \cite{Sev06,Dub11,Vos15}.

The description of weak decays using a model-independent
Effective-Field-Theory approach \cite{Cir10} has recently made
possible direct comparisons of sensitivity between searches for exotic interactions carried out
at low energies and at the Large Hadron Collider (LHC) \cite{Bha12,Cir13}.
Under the general assumption that the scale of new physics
occurs at much higher energies than those accessible at the LHC,
it appears that experiments at the LHC provide
tight constraints on scalar and tensor couplings involving right-handed
neutrinos \cite{Cir13}, which are more stringent than those obtained from
their quadratic contributions to $\beta$ decays. On the other hand, for interactions involving
left-handed neutrinos, experiments in nuclear and neutron
decay can potentially be competitive with constraints reached or to be
reached at the LHC provided they address observables
that are linear in the exotic couplings with sufficient precision~\cite{Bha12,Cir13}. 
This competition is also possible thanks to recent calculations of the
corresponding hadronic form factors, which are now known with $\sim 10\%$ precision~\cite{Bha12,Gon13,Bha16}.

One of the most sensitive observables to these nonstandard scalar
and tensor interactions in nuclear $\beta$ decay, which is linear
in the couplings and that currently provides the tightest
bounds on them \cite{Vos15,Nav13},
is the Fierz interference term. This term enters many measured
observables and affects in particular
the shape of the $\beta$ energy spectrum. The potential of
the Fierz term to probe the presence of exotic couplings has motivated
new measurements of $\beta$ energy
spectra with improved precision
\cite{Sev14}. In nuclear $\beta$ decay there is a variety of
transitions that can be selected for such measurements but
there appears to be some confusion as to which are the most
sensitive ones in terms of their kinematic signature to the Fierz
term. We first address this point below by discussing the sensitivity
to the Fierz term of the total decay rate, of the $\beta$ energy
spectrum and of the recoil momentum spectrum. We then point out the improper
use of some available experimental data of the $\beta-\nu$ angular correlation 
coefficient
and its relation to the Fierz term, 
and discuss the implications of those errors on the extraction of
constraints on exotic couplings.

%%%%%%%%%%%%%%%%%%%%%%%%%%%%%%%%%%%%%%%%%%%%%%%%%%%%%%%%%%%%%%%%%%%%%%%%%%%%%%%%
\section{Total decay rate}
\label{sec:totalrate}

We restrict ourselves to allowed $\beta$ decay
transitions described by
the statistical weight (phase space) of the form
\begin{equation}
P(W)dW = pWq^2 dW~,
\label{eq:phaseSpace}
\end{equation}
where $p$ and $W$ are respectively the momentum
and total energy of the $\beta$ particle
and $q = W_0-W$ is the momentum
of the neutrino, with $W_0$ the maximal
total energy of the $\beta$ particle.
We also consider two dynamic terms:
the Fierz term, $b$, and the $\beta-\nu$ angular correlation
coefficient, $a$. 
The decay rate function, averaged over
the spin variables of the nucleus and the electron,
is proportional to \cite{Kof54,Lee56}
\begin{equation}
N(W,\theta) dW d\Omega_\theta =
P(W) \left[1 + b\frac{m}{W} + a \frac{p}{W} \cos\theta \right] dW d\Omega_\theta~,
\label{eq:NEdEdO}
\end{equation}
where $m$ is the mass of the $\beta$ particle,
$\theta$ is the angle between the momentum directions of
the $\beta$ particle and the neutrino, and $d\Omega_\theta$ is
the differential solid angle around $\theta$.
For simplicity, we do not include contributions
due to recoil order terms \cite{Hol74}
or to Coulomb and radiative corrections \cite{Wil93a,Wil93b},
and neglect effects due to the neutrino mass.

\begin{figure}[!htb]
\centerline{
\includegraphics[width=0.80\linewidth]{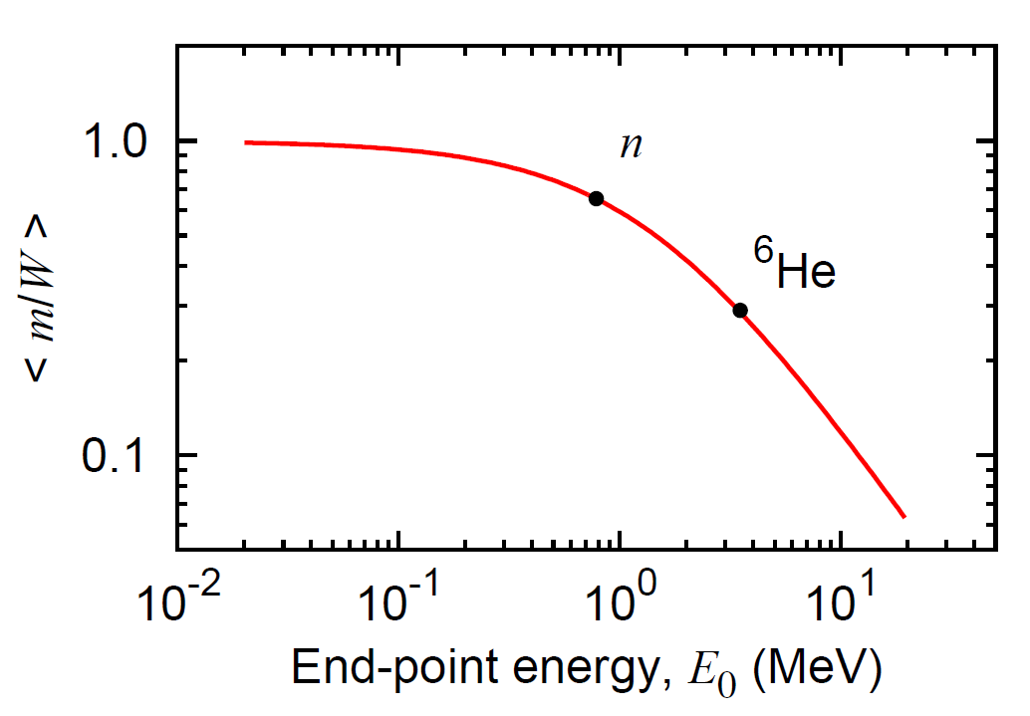}
}
\caption{(Color on-line) Variation of the sensitivity factor
$\langle m/W \rangle$ as a function of the endpoint energy $E_0$.}
\label{fig:avgdWm1}
\end{figure}

The integration of Eq.~(\ref{eq:NEdEdO}) over the kinematic variables
of the $\beta$ particle and the neutrino, normalized by the integral
over the phase space, gives
\begin{equation}
N_0 = 1 + b\langle \frac{m}{W} \rangle~,
\label{eq:N0}
\end{equation}
where $\langle m/W \rangle$ denotes the average of $m/W$ over the
statistical weight given by Eq.~(\ref{eq:phaseSpace}).
Figure \ref{fig:avgdWm1} shows the variation of the factor
$\langle m/W \rangle$ as a function of the endpoint
energy, $E_0 = W_0 - m$,
for values in the range 20~keV to 20~MeV.
For reference, the values for neutron decay ($E_0 = 782$~keV) and
for $^6$He decay ($E_0 = 3.50$~MeV) are indicated with black points.
It is obvious that $\langle m/W \rangle$ increases monotonically
toward lower endpoint energies and tends asymptotically to
1 since the kinetic energy
in the denominator becomes negligible relative to the electron
mass.

This property has been exploited to extract very stringent
constraints on scalar couplings from the contribution of the Fierz
term to the ${\cal F}t$-values in super-allowed pure Fermi
transitions \cite{Har15}. Nuclei with the lowest endpoint energies,
such as $^{10}$C and $^{14}$O, have the largest sensitivity to the
Fierz term, whereas the $b$-contamination to the ${\cal F}t$-values
of transitions with larger endpoints, such
as $^{26m}$Al, is smaller.

It is clear from Fig.~\ref{fig:avgdWm1}
that, from a purely statistical standpoint, the uncertainty on the Fierz
term extracted from a measurement of the rate in Eq.~(\ref{eq:N0})
would decrease monotonically toward lower energies. For a sample with
$10^8$ events, the smallest statistical uncertainty would
be $\Delta b = 10^{-4}$.

%%%%%%%%%%%%%%%%%%%%%%%%%%%%%%%%%%%%%%%%%%%%%%%%%%%%%%%%%%%%%%%%%%%%%%%%%%%%%%%%
\section{Differential distributions}
\label{sec:diffdistr}

The monotonic increase of sensitivity
to $b$ in Eq.~(\ref{eq:N0})
does not imply, however, that this property also holds
when the Fierz term is extracted
from the measurement of a differential distribution such as
the $\beta$ energy spectrum or the recoil momentum spectrum.
This is so, simply because in differential distributions one measures
the effect on the
shape of the distribution and not on the number of events.
To illustrate this quantitatively we have performed
simple Monte-Carlo studies
where the statistical uncertainty on the Fierz term is determined
from fits of differential spectra.

%%%%%%%%%%%%%%%%%%%%%%%%%%%%%%%%%%%%%%%%%%%%%%%%%%%%%%%%%%%%%%%%%%%%%%%%%%%%%%%%
\subsection{The $\beta$ energy spectrum}
\label{subsec:betaSpec}

We consider first the distribution
in electron energy, resulting from the integration of Eq.~(\ref{eq:NEdEdO})
over the directions of the neutrino,
\begin{eqnarray}
N_e(W) dW &\propto& P(W) \cdot \left(1 + b\,\frac{m}{W} \right) dW \label{eq:NedE1} \\
          &=& \left[ P(W) + b\,g(W) \right] dW~. \label{eq:NedE2} 
\end{eqnarray}
We have generated $\beta$-energy spectra following the shape of the phase
space $P(W)$ in Eq.~(\ref{eq:phaseSpace}), for different values
of the endpoint energy, $E_0$. Each spectrum contained $10^8$ events.
The generated spectra were then fitted between 5\% and 95\% of their kinetic 
energy range, with a function given by Eq.~(\ref{eq:NedE1}). 
The fits had two free parameters: the overall normalization
and the Fierz term $b$.

\begin{figure}[!htb]
\centerline{
\includegraphics[width=0.80\linewidth]{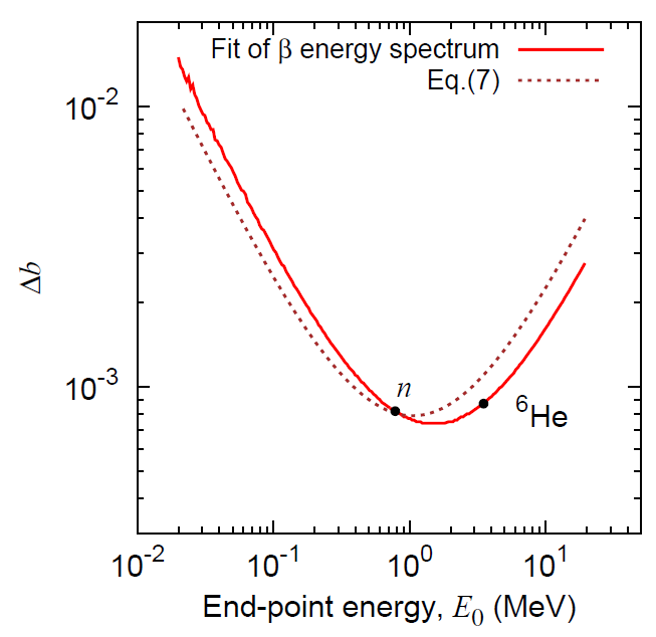}
}
\caption{(Color on-line) The solid red line shows the 1$\sigma$ statistical
uncertainties obtained
from fits of simulated $\beta$ energy spectra as a function of the endpoint
energy $E_0$. The dashed brown line shows the result obtained
with the  approximation given by Eq.~(\ref{eq:Delta_b_approx}).}
\label{fig:Db_betaEner}
\end{figure}

\begin{figure}[!htb]
\centerline{
\includegraphics[width=0.80\linewidth]{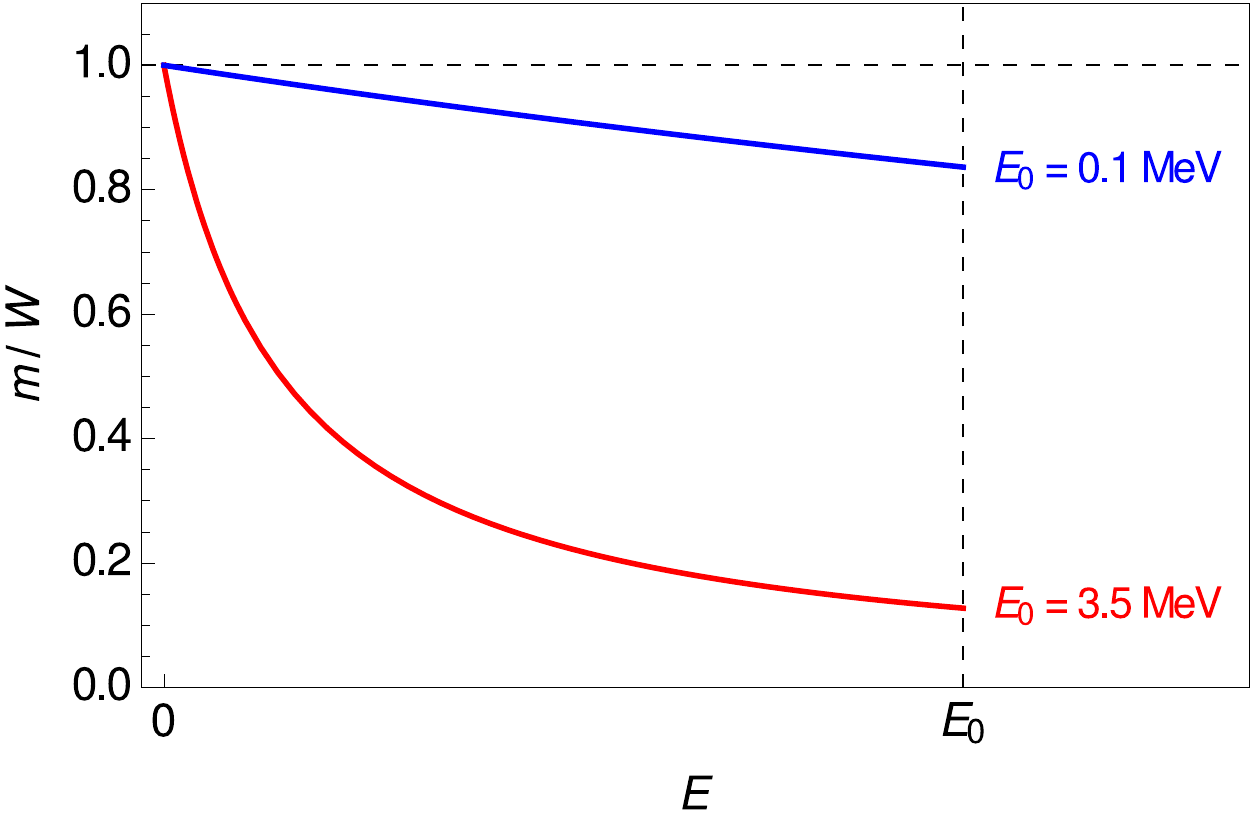}
}
\caption{(Color on-line) Energy dependence of the ratio $m/W$,
for two different endpoint energies.
Notice that the $E$-axis is rescaled to ease the comparison.}
\label{fig:moW}
\end{figure}

The red solid curve on 
Fig.~\ref{fig:Db_betaEner} shows the $1\sigma$ statistical uncertainty
on the Fierz term obtained from these fits as a function of the
endpoint energy, $E_0$.
For endpoint energies larger than about 1-2~MeV, the statistical
uncertainty increases roughly linearly with the endpoint energy,
due to the $1/W$ factor. For endpoint energies smaller than
1-2~MeV the statistical uncertainty does not decrease monotonically
but, instead, it also increases and equally fast on the log-log scale. 
The origin of this loss of sensitivity toward smaller endpoint energies is rather simple. The sensitivity to $b$ in the fits is driven by the differences in shape between $g(W)$ and $P(W)$ in Eq.~(\ref{eq:NedE2}).
As the average kinetic energy becomes smaller, these two functions become identical and the fitting function becomes proportional to $(1+b)$ with a loss of the specific kinematic signature to $b$.
In other words, although the effect of the $b$ term in the overall normalization, \textit{c.f.}
Eq.~(\ref{eq:N0}), is maximal for very low
endpoints, its effect on the shape becomes very weak
simply because the factor $m/W$ becomes
almost energy independent, namely $m/W\approx 1$. This is
illustrated in Fig.~\ref{fig:moW}.

It is interesting to note that neutron decay and $^6$He decay,
which are being considered for precision measurements
of the $\beta$ energy spectrum
\cite{Poc09,Huy16}, have very comparable kinematic
sensitivities (Fig.~\ref{fig:Db_betaEner}).
It is worth stressing that transitions with endpoint energies
in the range $E_0$~=~200-300~keV, such as $^{45}$Ca, which are also
currently the focus of new projects \cite{Sev14}, are a
factor of 2 less sensitive that those with $E_0$~=~1-2~MeV.
Last but not least, it is observed 
that transitions with endpoint energies in the range
0.6-3.8~MeV have sensitivities that are within 20\% of the
optimal kinematic sensitivity and are therefore ideal to search for
a non-zero Fierz term, at least on pure statistical grounds.

The observation of the loss of sensitivity toward smaller endpoint
energies obtained in the Monte-Carlo study can be derived from simple arguments.
For this we consider the central region of a $\beta$ energy spectrum.
If we divide the kinetic energy range in four equal intervals, the
central values of the kinetic energies of the second and third intervals
are respectively $E_2 = 3E_0/8$ and $E_3 = 5E_0/8$.
Following Eq.~(\ref{eq:NedE1}), the ratio between the number of events
in these two central bins is approximately proportional to 
\begin{equation}
1 + b\cdot \left(\frac{m}{m+E_2} - \frac{m}{m+E_3} \right)~.
\label{eq:b_approx} 
\end{equation}
We further assume
that, for a given spectrum, this ratio dominates the sensitivity to $b$.
With $N$ events in the spectrum, the statistical uncertainty on the Fierz term
extracted from such a ratio will approximately be determined from
\begin{equation}
\Delta b \cdot \left(\frac{m}{m+E_2} - \frac{m}{m+E_3} \right) \approx
\frac{1}{\sqrt{N}}~.
\label{eq:Delta_b_approx}
\end{equation}
The values of $\Delta b$ extracted from Eq.~(\ref{eq:Delta_b_approx}),
for $N = 10^8$, are shown by
the dashed brown line in Fig.~\ref{fig:Db_betaEner} as a function of the
endpoint energy. This crude 
approximation gives the main trend of the sensitivity, independently of
any reference to a fit. As the endpoint
energy becomes smaller, the difference in Eq.~(\ref{eq:Delta_b_approx}) also
becomes smaller resulting in the increase of $\Delta b$.

%%%%%%%%%%%%%%%%%%%%%%%%%%%%%%%%%%%%%%%%%%%%%%%%%%%%%%%%%%%%%%%%%%%%%%%%%%%%%%%%
\subsection{The recoil momentum spectrum}
\label{subsec:recoilSpec}

As a second example, we consider the extraction of $b$ from the 
distribution of the recoil nucleus momentum $r$. The transformation of
Eq.~(\ref{eq:NEdEdO}) in terms of the
recoil momentum gives the distribution~\cite{Kof54,Kof48}
\begin{equation}
N(W,r) dW dr = \frac{1}{2} r W q \left[1 + b \frac{m}{W}
            + a \frac{r^2 - p^2 - q^2}{2Wq} \right] dW dr~, \label{eq:NdEdr}
\end{equation}
where, for a given value of $r$ the total energy of the $\beta$ particle
varies between 
\begin{eqnarray}
W_{\rm min} (r) &=& \frac{(W_0 - r)^2 + m^2}{2(W_0 - r)} \label{eq:Emin}~\,,\\
W_{\rm max} (r) &=& \frac{(W_0 + r)^2 + m^2}{2(W_0 + r)} \label{eq:Emax}~\,.
\end{eqnarray}
The integration of Eq.~(\ref{eq:NdEdr}) over the $\beta$ energy gives the
recoil momentum distribution
\begin{equation}
Q(r)dr = \frac{r}{12} \left[ Q_0(r) + b\, Q_1(r) + a\, Q_2(r) \right]dr~,
\label{eq:Qrdr}
\end{equation}
where
\begin{equation}
Q_i(r) = \left[ W F_i(W, r) \right]^{W_{\rm max}(r)}_{W_{\rm min}(r)},
\label{eq:Qi}
\end{equation}
and
\begin{eqnarray}
F_0(W,r) &=& W(3W_0-2W) \label{eq:F0Wr} ~,\\
F_1(W,r) &=& 3 m (2W_0-W) \label{eq:F1Wr} ~,\\
F_2(W,r) &=& 3 \left( r^2 + m^2 - W_0^2 + W_0 W - \frac{2}{3}W^2 \right). \label{eq:F2Wr}
\end{eqnarray}
In order to determine the sensitivity to the Fierz term from a measurement of
a recoil momentum spectrum we have again used a Monte-Carlo method. We
have generated distributions following
Eq.~(\ref{eq:Qrdr}), with $b = 0$, for pure Fermi ($a = a_F = 1$) and pure
Gamow-Teller ($a = a_{GT} = -1/3$) transitions, and for different values of
the endpoint energy. Each spectrum contained $10^8$ events.  The spectra
were then fitted between 5\% and 95\% of the momentum range, with a function
given also by Eq.~(\ref{eq:Qrdr}), with two free parameters: the overall
normalization and $b$.
Since the $a$ coefficient depends on the nonstandard couplings but quadratically,
the procedure above is equivalent to neglecting those contributions.

\begin{figure}[!hbt]
\centerline{
\includegraphics[width=0.90\linewidth]{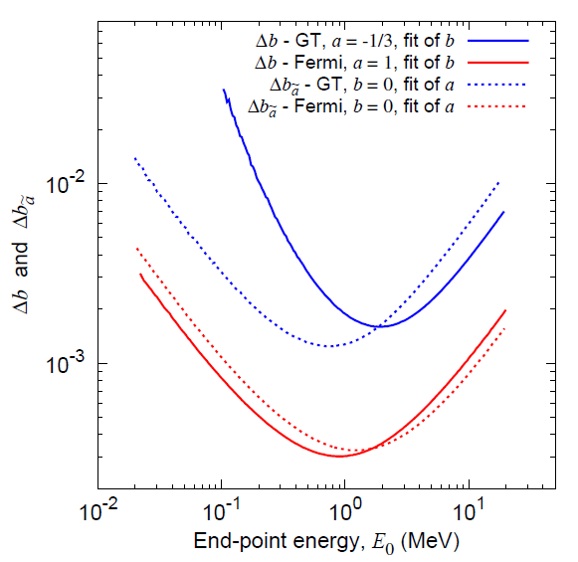}
}
\caption{(Color on-line) The 1$\sigma$ statistical uncertainties on $b$ obtained
from fits of simulated recoil momentum spectra as a function of the endpoint
energy $E_0$. The solid lines represent $\Delta b$ and correspond to Fermi
(red) and Gamow-Teller (blue) transitions for which the value of $a$ was
respectively fixed to $a_F$ or $a_{GT}$. The dotted lines represent $\Delta b_{\tilde{a}}$,
that is the uncertainty on $b$ obtained
by incorrectly re-interpreting a fit of $a$ with $b = 0$,
through the $\tilde{a}$-prescription, \textit{c.f.} Eq.~(\ref{eq:Delta_ba}).}  
\label{fig:Db_recoilMom}
\end{figure}

The solid lines in Fig.~\ref{fig:Db_recoilMom} show the $1\sigma$ statistical
uncertainty on the Fierz term obtained from these fits as a function of the
endpoint energy for pure Fermi (red curve) and pure Gamow-Teller (blue curve)
transitions. For both types of transitions, the dependence is
qualitatively the same as for the extractions of $b$ from the $\beta$ energy
spectrum, and has the same origin. This behavior can again
be understood analytically from simple arguments like those used to
deduce Eq.~(\ref{eq:Delta_b_approx}), but the expressions are lengthier
in this case. Figure \ref{fig:Db_recoilMom} shows that
the presence of the term proportional to $a$ in Eq.~(\ref{eq:Qrdr}) gives rise
to different sensitivities for Fermi and Gamow-Teller transitions.
 
Obtaining an analogous sensitivity curve for mixed transitions is not possible since
in these decays the correlation coefficient $a$ is a function
of the Gamow-Teller to Fermi mixing ratio, $a(\rho)$, and can therefore not be
fixed{\footnote{This also means that one can extract both $b$ and $\rho$ (or equivalently $b$ and $a$) simultaneously
from the recoil spectrum.}}.
More precisely, for pure Fermi and Gamow-Teller transitions, the value of $a$ is
essentially determined by angular momentum arguments and can then be fixed to the
SM value. In contrast, for mixed transitions, the expression of $a$ depends on $\rho$
that, in turn, is extracted from the measurement of another correlation coefficient
such as $\tilde{A}$, or from the comparison between the $ft$-value of the mixed
transition and ${\cal F}t(0^+\rightarrow 0^+)$.
However, these observables also receive a contamination due to $b$ that affects
the $\rho$ extraction and that propagates to $a$.
This ``indirect" effect of the Fierz term, some times ignored in the literature (see e.g.
Ref.~\cite{Vet08}), is of the same order as the direct effect in the recoil
spectrum, and therefore it has to be taken into account in the analysis, since
it can lead to strong suppressions of sensitivity.

%%%%%%%%%%%%%%%%%%%%%%%%%%%%%%%%%%%%%%%%%%%%%%%%%%%%%%%%%%%%%%%%%%%%%%%%%%%%%%%%
\section{Correlation coefficients}
\label{sec:corrCoeff}

Many past experiments extracted values of various correlation
coefficients within an analysis that sets the Fierz term to zero.
There are different reasons to do this.
For instance, the extraction of the axial-vector coupling $g_A$ from
measurements of the asymmetry parameter $A$ in neutron decay~\cite{Mun13,Men13},
is usually performed within the SM framework so that nonstandard effects are
simply ignored.
Other experiments analyzed the data with the Fierz term set to zero,
because existing bounds on it were strong enough, and they focused on
the sensitivity to %nonstandard 
interactions entering
mainly through quadratic contributions, as e.g. in Ref.~\cite{Joh63}.

It was noticed long ago \cite{Pau70} that, if a correlation coefficient $X$
has been extracted from measurements of an asymmetry in counting rates
assuming $b=0$, it
can be easily re-interpreted by including a non-zero Fierz term through
the expression
\begin{equation}
\tilde{X} = \frac{X} {1+ b \langle m/W \rangle^\prime} \,.
\label{eq:Xtilde}
\end{equation}
The factor $\langle m/W \rangle^\prime$ denotes here the integration over the
measured interval of the $\beta$ energy spectrum.
Such an expression was introduced for ``standard'' experimental determinations of parameters like $A$, $B$, $G$
and $D$ \cite{Jac57a,Pau70}.
This prescription has been used in global analyses of data in neutron
and nuclear decays \cite{Sev06,Kon10,Wau14,Vos15} to take into account the
various measurements where the Fierz term was not included in the original
analysis.

Operationally, Eq.~(\ref{eq:Xtilde}) was noticed to be valid when
the measured correlation coefficient is deduced from an asymmetry resulting
from the sign inversion of some kinematic variable or the
inversion of the analysis direction of a variable, such as the direction of a magnetic field. 
Formally, the prescription is valid when the integration limits of the $\beta$-particle energy, $W$,
and the relevant kinematic variable, $\theta$, are independent, and then the $W$ integration can be performed without introducing a genuine $\theta$-dependence in the Fierz term. More precisely one would have\newpage
%\begin{samepage}
%
\begin{eqnarray}
%N(W,\theta) dW d\theta &=& G(\theta)H(W) \left[ 1+ b \, \frac{m}{W} + X\,R(W,\theta) \right] dW d\theta \nonumber\\
N(W,\theta) dW d\theta 
&=& G(\theta)H(W) 
\\&& \times \left[ 1+ b \, \frac{m}{W} + X\,R(W,\theta) \right] dW d\theta \nonumber\\
%\to N(\theta) d\theta &=& G(\theta)\langle H(W) \rangle \left( 1+ b \, \langle \frac{m}{W} \rangle + X\,\langle R(W,\theta) \rangle \right)\,.\\
%\to N(\theta) d\theta &\sim& G(\theta) \left( 1 + \tilde{X}\,R(\theta) \right)\,,\\
\to N(\theta) d\theta &\sim& G(\theta) \left( 1+ b \, \langle \frac{m}{W} \rangle \right) \left[ 1 + \tilde{X}\,R(\theta) \right] d\theta,\nonumber
\label{eq:gg}
\end{eqnarray}
%\end{samepage}
%
where $R(\theta)\equiv \langle R(W,\theta) \rangle_W$, and where we have also
assumed that the $W$- and $\theta$-dependences of the normalization function are factorizable. It is easy to see from this expression that the $\tilde{X}$-prescription will then apply not only for standard asymmetries, but also for measurements of the $\theta$ differential distribution \cite{Chi84}.

However, the recoil momentum $r$ and the $\beta$-particle energy $W$ are not independent.
As a result, the term $Q_1(r)$ in Eq.~(\ref{eq:Qrdr})
has a different dependence on the recoil momentum than the term $Q_0(r)$ and cannot be factorized
to produce a term of the form $(1 + b\langle m/W \rangle)$.
This is particularly relevant for measurements of the $\beta-\nu$ angular
correlation coefficient. Unless this coefficient is extracted from an asymmetry
in decay rates \cite{Wie09} or for a fixed $\beta$ energy \cite{Gri68}, the measured
distribution will contain both $b$ and $a$ terms with different recoil
momentum dependence. 

Consequently, it is improper to use the prescription given in
Eq.~(\ref{eq:Xtilde}) to re-interpret previous
extractions of the $\beta-\nu$ angular correlation
coefficient where $b$ was set to zero and only $a$ was fitted. 
For the most precise measurement in a Gamow-Teller decay \cite{Joh63}, $a$ was actually 
extracted from a differential measurement of the recoil momentum distribution,
as given by Eq.~(\ref{eq:Qrdr}). 
The result was originally used to constrain possible tensor couplings through their
quadratic contribution to $a$, assuming $b=0$ and assigning an
uncertainty $\Delta b = 0.012$ on the basis of a previous measurement in $^{22}$Na.
This uncertainty on $b$ is furthermore included in the total uncertainty of
the quoted value of $a$.
This measurement of $a$ in $^6$He, with a later revision \cite{Glu98}, has
been used in several global fits \cite{Sev06,Wau14,Vos15}, reviews \cite{Sev11,Nav13}
and articles~\cite{Vet08,Pit08}, where $a$ has been reinterpreted as $\tilde{a}$.

In order to illustrate the impact on the extraction of $b$ from this incorrect interpretation
when applied to allowed pure Fermi and Gamow-Teller transitions, we have performed 
additional fits of the recoil momentum spectra, this time with $b=0$ and $a$
left as free parameter. 
If the fitted value of $a$ is reinterpreted as $\tilde{a}$ one can extract the uncertainty on the Fierz term using Eq.~(\ref{eq:Xtilde}),
\begin{equation}
\Delta b_{\tilde{a}} = \frac{\Delta \tilde{a}}{|a_{SM}|}  \,\langle \frac{m}{W} \rangle^{-1} \,.
\label{eq:Delta_ba}
\end{equation}

The dotted lines in Fig.~\ref{fig:Db_recoilMom} show the $1\sigma$ statistical
uncertainty
on the Fierz term obtained from these fits as a function of the
endpoint energy for pure Fermi (red curve) and pure Gamow-Teller (blue curve)
transitions. The dependence on the endpoint energy is again qualitatively
similar to the direct extraction of $b$, with a loss of sensitivity toward
small endpoint values. 
This is so because the loss of sensitivity in the extraction of $a$ from the
differential distribution
dominates over the mild increase of the factor $\langle m/W \rangle$. 
For Fermi transitions the relative differences between the two curves are small,
of about 25\% at small endpoint values and of 20\% at high values.
However, for Gamow-Teller transitions the differences between the two
results diverge for low endpoint values. It can be shown analytically that
\begin{equation}
\lim_{E_0 \to 0} \frac{\Delta b}{\Delta b_{\tilde{a}}} \approx
\left| \frac{3a}{1 + 3a} \right|~,
\label{eq:lim_0}
\end{equation} 
where $\Delta b$ is the uncertainty extracted from direct fits of $b$.
This limit tends to infinity for $a = a_{GT}$ and explains the divergence
observed between the values of $\Delta b$ and $\Delta b_{\tilde{a}}$.
If the $\tilde{a}$-prescription would have been applied to a
Gamow-Teller transition with an end-point of $E_0 = 100$~keV the error on
the uncertainty of $b$ would have been of about an order of magnitude.
From this analysis, the relative difference of the two Gamow-Teller
curves at the $^6$He endpoint, $E_0 = 3.5$~MeV, is 30\%. The position
between the two
curves, to the right of the intersection point, indicates that
the values of the Fierz term extracted from an incorrect reinterpretation
of $a$ are less precise than the values extracted from a
direct fit of $b$ using the same data.
This result should however be taken with the proper caution,
as the above simplified analysis neglects systematic effects and other
details in the data analysis.

\begin{figure}[!hbt]
\centerline{
\includegraphics[width=0.90\linewidth]{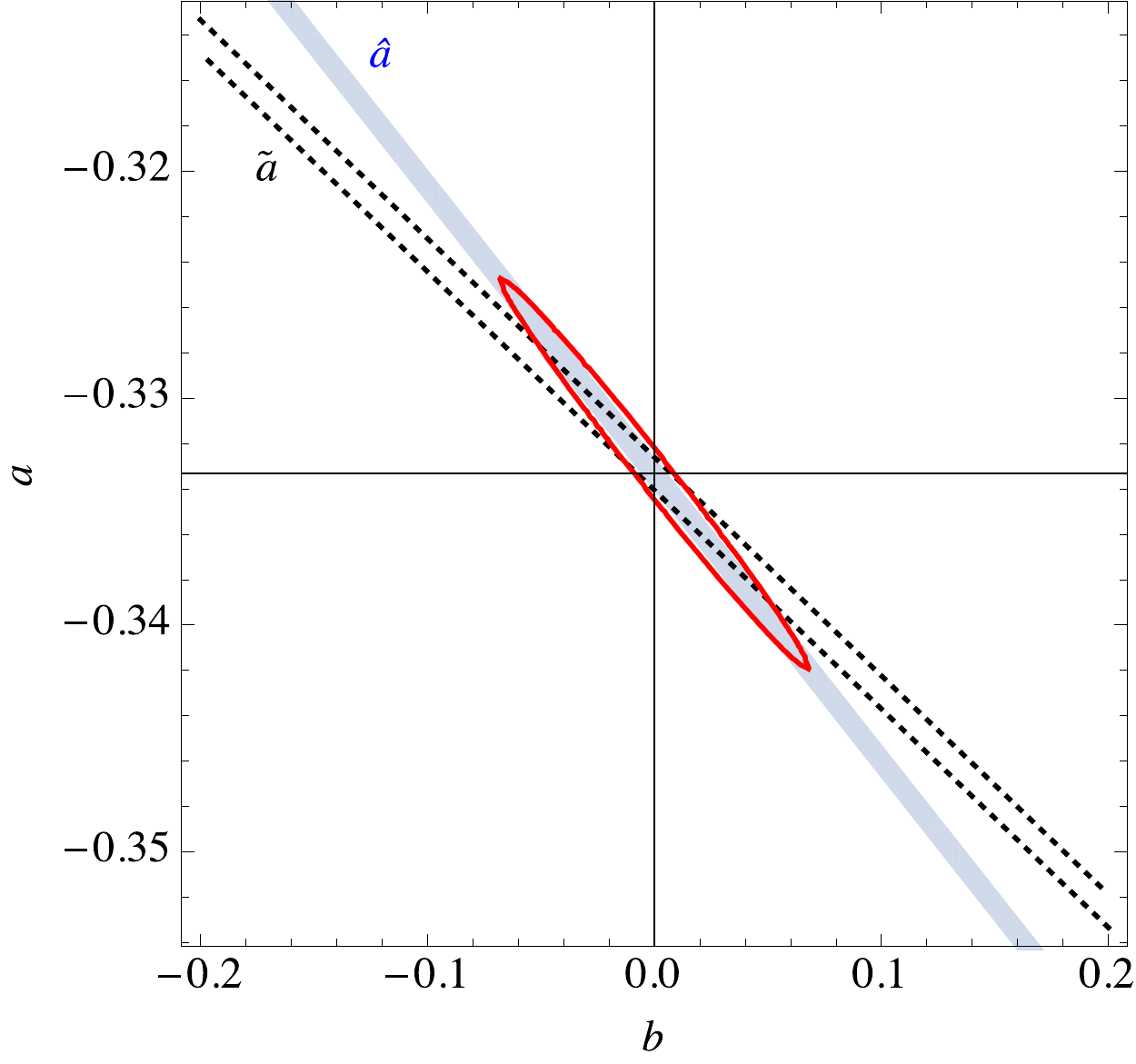}
}
\caption{(Color on-line)
The solid red ellipse shows the 1$\sigma$ region obtained from a fit of simulated
recoil momentum spectra with $10^7$ events, for the $^6$He
decay, where both $a$ and $b$ were left as free parameters.
The blue filled band shows the 1$\sigma$ bound on the combination $\hat{a}=a+0.127\,b$,
whereas the black dotted lines represent the 1$\sigma$ bound obtained using the $\tilde{a}$-prescription. }  
\label{fig:ahat}
\end{figure}

A somewhat different way of analyzing the error of the $\tilde{a}$-prescription
is obtained by performing a fit of the differential recoil
distributions with both $a$ and $b$ as free parameters.
Fig.~\ref{fig:ahat} shows the
result we obtained for the $^6$He decay with $10^7$ events in the
spectrum. One observes that there is indeed a large correlation between $a$ and $b$,
i.e. that a certain linear combination of them, namely $\hat{a}=a+0.127\,b$,
is strongly constrained. We see, however, that the $\hat{a}$-band is not aligned with
the one obtained using
$\tilde{a}\approx a (1- \langle m/W \rangle\, b) \approx a(1 - 0.286\,b)$.

Concerning pure Fermi transitions, the two most precise extractions of $a$ are those
in $^{32}$Ar~\cite{Ade99} and in $^{38m}$K decay~\cite{Gor05}, which were also used
in global fits \cite{Sev06,Vos15}. 
The direct observable was the delayed proton spectrum
following $^{32}$Ar decay and the time-of-flight spectra of $^{38}$Ar$^{n+}$ ions in $^{38m}$K decay,
where the $\beta$ particle spectrum was either totally or partly integrated.
Thus, according to the results presented above, the $\tilde{a}$-prescription is not applicable in such
measurements.
Although the corresponding analyses were performed using a parameter called
$\tilde{a}$~\cite{Ade99,Gor05,Gor08}, it is important to notice that this parameter 
does not follow the standard definition of $\tilde{a}$,
\emph{c.f.} Eq.~(\ref{eq:Xtilde}), also used in the present work.
Instead, it corresponds to what we called $\hat{a}$ above, i.e. a linear
combination of $a$ and $b$ that is strongly constrained by the fit,
with coefficients that
have to be calculated for each transition \emph{a posteriori}~\cite{Gar16,Gor16}.
For example, in the $^{32}$Ar experiment the value of the coefficient that multiplies
$b$ in $\hat{a}$, namely 0.1913 \cite{Ade99}, is 11\% smaller than the value
of $\langle m/W \rangle = 0.214$ expected for this transition in the $\tilde{a}$
expression~\cite{Ade99,Gar16}.
As can be inferred from Fig.~\ref{fig:Db_recoilMom}, the numerical expressions
of $\tilde{a}$ and $\hat{a}$ are quite similar for pure Fermi
transitions independently of the endpoint energy.  It is however important
to notice that conceptually these two parameters are completely
different.

%%%%%%%%%%%%%%%%%%%%%%%%%%%%%%%%%%%%%%%%%%%%%%%%%%%%%%%%%%%%%%%%%%%%%%%%%%%%%%%%
\section{Concluding remarks}

The Fierz term, $b$, is one of the few parameters
in $\beta$ decays that is linearly sensitive to nonstandard interactions
and its precise measurement represents a competitive New-Physics probe even in the LHC era~\cite{Bha12,Cir13,Nav13}. In this work we have analyzed a few aspects that are relevant for (i) the selection of sensitive nuclear decays for future experiments, and (ii) the
extraction of precise and correct bounds on $b$ from past and future measurements.

In Sec.~\ref{sec:totalrate}-\ref{sec:diffdistr} we have analyzed how the statistical sensitivity to the Fierz term changes with the endpoint energy of the decay. We showed that although the effect in the overall normalization is maximal for very low endpoints, its effect on the $\beta$-energy
and recoil momentum differential distribution goes to zero in that limit. For each case we identified the window of endpoint values where the sensitivity to $b$ is maximal. These results are relevant for the selection of the most sensitive transitions in measurements of $b$. It is important to stress that the
kinematic sensitivity is only one, and possibly the simplest, among several criteria for such a selection.
Other criteria for measurements of $\beta$ spectrum shapes are related with the size and accuracy of:
1) Coulomb and radiative corrections to the $\beta$ spectrum \cite{Wil93a,Wil93b};
2) the form factors which enter the weak hadronic currents in recoil terms \cite{Hol74}; and
3) instrumental effects such as the scattering of $\beta$ particles in matter. For example, atomic effects due to screening are known to be large for low energy $\beta$ particles from transitions in medium and heavy mass nuclei \cite{Mou14} and require therefore particular
attention in precision measurements. On the instrumental side, the effects of scattering of electrons from matter in the setup and their back-scattering
from detector surfaces also increases toward lower energies and the description of
the processes with current simulation tools \cite{Gol08,Sot13} is not yet sufficiently
accurate for competitive measurements of
the $\beta$ spectrum shape.
These two criteria tend also to disfavor transitions with small endpoint energies in the selection of candidates.

In Sec.~\ref{sec:corrCoeff} we discussed what we called the $\tilde{X}$-prescription and its relation to
the Fierz term.
We have shown explicitly that the prescription cannot be applied to values of $a$ extracted from
differential measurements of the recoil momentum distributions and we have explained under which conditions the
procedure is justified.
The prescription has been applied however, in a somewhat undiscriminated way,
in several recent global fits \cite{Sev06,Kon10,Wau14,Vos15}
for the re-interpretation of values of $a$ extracted in $^6$He and in neutron decays.
The numerical impact of this misinterpretation on the constraints of exotic couplings
extracted in global fits is in most cases quite small, simply because
the precision achieved so far in measurements of recoil distributions is moderate. The
associated constraints are therefore not competitive with determinations of $b$ from
other observables that dominate the fits.
However it is important to notice that this might change in the near future
with new generation measurements of the recoil spectrum \cite{Kne13}.
 
Finally, in most of the numerical analyses performed in this work we neglected quadratic
nonstandard effects that contribute to $a$. We would like to emphasize that it would be
suitable that future analyses of new measurements of differential recoil
distributions fit both $a$ and $b$ simultaneously, and provide their correlation,
as in Fig.~\ref{fig:ahat}.
One could then study specific cases with left-handed and right-handed couplings separately.
For instance, for exotic interactions involving right-handed neutrinos, the
linear terms are absent and then the quadratic terms become the leading ones.

%
%%%%%%%%%%%%%%%%%%%%%%%%%%%%%%%%%%%%%%%%%%%%%%%%%%%%%%%%%%%%%%%%%%%%%%%%%%%%%%%%
\begin{acknowledgments}
We thank J.~Behr, A.~Garc\'{i}a, A.~Gorelov, X.~Fl\'echard, K.~Minamisono, F.~Wauters and A.R.~Young for clarifications and fruitful discussions.
This work was supported in part by the U.S. National Science Foundation under grant number PHY-11-02511.
M.G.-A. is grateful to the LABEX Lyon Institute of Origins (ANR-10-LABX-0066) of the Universit\'e de Lyon for its financial support within the program ANR-11-IDEX-0007 of the French government.
\end{acknowledgments}
%%%%%%%%%%%%%%%%%%%%%%%%%%%%%%%%%%%%%%%%%%%%%%%%%%%%%%%%%%%%%%%%%%%%%%%%%%%%%%%%
\bibliography{thebibliography}

\begin{thebibliography}{42}
\expandafter\ifx\csname natexlab\endcsname\relax\def\natexlab#1{#1}\fi
\expandafter\ifx\csname bibnamefont\endcsname\relax
  \def\bibnamefont#1{#1}\fi
\expandafter\ifx\csname bibfnamefont\endcsname\relax
  \def\bibfnamefont#1{#1}\fi
\expandafter\ifx\csname citenamefont\endcsname\relax
  \def\citenamefont#1{#1}\fi
\expandafter\ifx\csname url\endcsname\relax
  \def\url#1{\texttt{#1}}\fi
\expandafter\ifx\csname urlprefix\endcsname\relax\def\urlprefix{URL }\fi
\providecommand{\bibinfo}[2]{#2}
\providecommand{\eprint}[2][]{\url{#2}}

\bibitem[{\citenamefont{Severijns et~al.}(2006)\citenamefont{Severijns, Beck,
  and Naviliat-Cuncic}}]{Sev06}
\bibinfo{author}{\bibfnamefont{N.}~\bibnamefont{Severijns}},
  \bibinfo{author}{\bibfnamefont{M.}~\bibnamefont{Beck}}, \bibnamefont{and}
  \bibinfo{author}{\bibfnamefont{O.}~\bibnamefont{Naviliat-Cuncic}},
  \bibinfo{journal}{Rev. Mod. Phys.} \textbf{\bibinfo{volume}{78}},
  \bibinfo{pages}{991} (\bibinfo{year}{2006}).

\bibitem[{\citenamefont{Dubbers and Schmidt}(2011)}]{Dub11}
\bibinfo{author}{\bibfnamefont{D.}~\bibnamefont{Dubbers}} \bibnamefont{and}
  \bibinfo{author}{\bibfnamefont{M.~G.} \bibnamefont{Schmidt}},
  \bibinfo{journal}{Rev. Mod. Phys.} \textbf{\bibinfo{volume}{83}},
  \bibinfo{pages}{1111} (\bibinfo{year}{2011}).

\bibitem[{\citenamefont{Vos et~al.}(2015)\citenamefont{Vos, Wilschut, and
  Timmermans}}]{Vos15}
\bibinfo{author}{\bibfnamefont{K.~K.} \bibnamefont{Vos}},
  \bibinfo{author}{\bibfnamefont{H.~W.} \bibnamefont{Wilschut}},
  \bibnamefont{and} \bibinfo{author}{\bibfnamefont{R.~G.~E.}
  \bibnamefont{Timmermans}}, \bibinfo{journal}{Rev. Mod. Phys.}
  \textbf{\bibinfo{volume}{87}}, \bibinfo{pages}{1483} (\bibinfo{year}{2015}).

\bibitem[{\citenamefont{Cirigliano et~al.}(2010)\citenamefont{Cirigliano,
  Jenkins, and Gonzalez-Alonso}}]{Cir10}
\bibinfo{author}{\bibfnamefont{V.}~\bibnamefont{Cirigliano}},
  \bibinfo{author}{\bibfnamefont{J.}~\bibnamefont{Jenkins}}, \bibnamefont{and}
  \bibinfo{author}{\bibfnamefont{M.}~\bibnamefont{Gonzalez-Alonso}},
  \bibinfo{journal}{Nucl. Phys. B.} \textbf{\bibinfo{volume}{830}},
  \bibinfo{pages}{95} (\bibinfo{year}{2010}).

\bibitem[{\citenamefont{Bhattacharya et~al.}(2012)\citenamefont{Bhattacharya,
  Cirigliano, Cohen, Filipuzzi, Gonz\'alez-Alonso, Graesser, Gupta, and
  Lin}}]{Bha12}
\bibinfo{author}{\bibfnamefont{T.}~\bibnamefont{Bhattacharya}},
  \bibinfo{author}{\bibfnamefont{V.}~\bibnamefont{Cirigliano}},
  \bibinfo{author}{\bibfnamefont{S.~D.} \bibnamefont{Cohen}},
  \bibinfo{author}{\bibfnamefont{A.}~\bibnamefont{Filipuzzi}},
  \bibinfo{author}{\bibfnamefont{M.}~\bibnamefont{Gonz\'alez-Alonso}},
  \bibinfo{author}{\bibfnamefont{M.~L.} \bibnamefont{Graesser}},
  \bibinfo{author}{\bibfnamefont{R.}~\bibnamefont{Gupta}}, \bibnamefont{and}
  \bibinfo{author}{\bibfnamefont{H.-W.} \bibnamefont{Lin}},
  \bibinfo{journal}{Phys. Rev. D} \textbf{\bibinfo{volume}{85}},
  \bibinfo{pages}{054512} (\bibinfo{year}{2012}).

\bibitem[{\citenamefont{Cirigliano et~al.}(2013)\citenamefont{Cirigliano,
  Gonz{\'a}lez-Alonso, and Graesser}}]{Cir13}
\bibinfo{author}{\bibfnamefont{V.}~\bibnamefont{Cirigliano}},
  \bibinfo{author}{\bibfnamefont{M.}~\bibnamefont{Gonz{\'a}lez-Alonso}},
  \bibnamefont{and} \bibinfo{author}{\bibfnamefont{M.~L.}
  \bibnamefont{Graesser}}, \bibinfo{journal}{J. High Energy Phys.}
  \textbf{\bibinfo{volume}{2013}}, \bibinfo{pages}{1} (\bibinfo{year}{2013}).

\bibitem[{\citenamefont{Gonz\'alez-Alonso and Martin~Camalich}(2014)}]{Gon13}
\bibinfo{author}{\bibfnamefont{M.}~\bibnamefont{Gonz\'alez-Alonso}}
  \bibnamefont{and}
  \bibinfo{author}{\bibfnamefont{J.}~\bibnamefont{Martin~Camalich}},
  \bibinfo{journal}{Phys. Rev. Lett.} \textbf{\bibinfo{volume}{112}},
  \bibinfo{pages}{042501} (\bibinfo{year}{2014}).

\bibitem[{\citenamefont{Bhattacharya et~al.}(2016)\citenamefont{Bhattacharya,
  Cirigliano, Cohen, Gupta, Lin, and Yoon}}]{Bha16}
\bibinfo{author}{\bibfnamefont{T.}~\bibnamefont{Bhattacharya}},
  \bibinfo{author}{\bibfnamefont{V.}~\bibnamefont{Cirigliano}},
  \bibinfo{author}{\bibfnamefont{S.}~\bibnamefont{Cohen}},
  \bibinfo{author}{\bibfnamefont{R.}~\bibnamefont{Gupta}},
  \bibinfo{author}{\bibfnamefont{H.-W.} \bibnamefont{Lin}}, \bibnamefont{and}
  \bibinfo{author}{\bibfnamefont{B.}~\bibnamefont{Yoon}}
  (\bibinfo{year}{2016}), \eprint{1606.07049}.

\bibitem[{\citenamefont{Naviliat-Cuncic and Gonz\'alez-Alonso}(2013)}]{Nav13}
\bibinfo{author}{\bibfnamefont{O.}~\bibnamefont{Naviliat-Cuncic}}
  \bibnamefont{and}
  \bibinfo{author}{\bibfnamefont{M.}~\bibnamefont{Gonz\'alez-Alonso}},
  \bibinfo{journal}{Ann. Phys. (Berlin)} \textbf{\bibinfo{volume}{525}},
  \bibinfo{pages}{600} (\bibinfo{year}{2013}).

\bibitem[{\citenamefont{Severijns}(2014)}]{Sev14}
\bibinfo{author}{\bibfnamefont{N.}~\bibnamefont{Severijns}},
  \bibinfo{journal}{J. Phys. G: Nucl. Part. Phys.}
  \textbf{\bibinfo{volume}{41}}, \bibinfo{pages}{114006}
  (\bibinfo{year}{2014}).

\bibitem[{\citenamefont{Kofoed-Hansen}(1954)}]{Kof54}
\bibinfo{author}{\bibfnamefont{O.}~\bibnamefont{Kofoed-Hansen}},
  \bibinfo{journal}{Dan. Mat. Fys. Medd.} \textbf{\bibinfo{volume}{28}},
  \bibinfo{pages}{1} (\bibinfo{year}{1954}).

\bibitem[{\citenamefont{Lee and Yang}(1956)}]{Lee56}
\bibinfo{author}{\bibfnamefont{T.~D.} \bibnamefont{Lee}} \bibnamefont{and}
  \bibinfo{author}{\bibfnamefont{C.~N.} \bibnamefont{Yang}},
  \bibinfo{journal}{Phys. Rev.} \textbf{\bibinfo{volume}{104}},
  \bibinfo{pages}{254} (\bibinfo{year}{1956}).

\bibitem[{\citenamefont{Holstein}(1974)}]{Hol74}
\bibinfo{author}{\bibfnamefont{B.~R.} \bibnamefont{Holstein}},
  \bibinfo{journal}{Rev. Mod. Phys.} \textbf{\bibinfo{volume}{46}},
  \bibinfo{pages}{789} (\bibinfo{year}{1974}), \bibinfo{note}{[Erratum: Rev.
  Mod. Phys. {\bf 48}, 673 (1976)]}.

\bibitem[{\citenamefont{Wilkinson}(1993{\natexlab{a}})}]{Wil93a}
\bibinfo{author}{\bibfnamefont{D.}~\bibnamefont{Wilkinson}},
  \bibinfo{journal}{Nucl. Instr. Meth. Phys. Res. A}
  \textbf{\bibinfo{volume}{335}}, \bibinfo{pages}{182 }
  (\bibinfo{year}{1993}{\natexlab{a}}).

\bibitem[{\citenamefont{Wilkinson}(1993{\natexlab{b}})}]{Wil93b}
\bibinfo{author}{\bibfnamefont{D.}~\bibnamefont{Wilkinson}},
  \bibinfo{journal}{Nucl. Instr. Meth. Phys. Res. A}
  \textbf{\bibinfo{volume}{335}}, \bibinfo{pages}{201 }
  (\bibinfo{year}{1993}{\natexlab{b}}).

\bibitem[{\citenamefont{Hardy and Towner}(2015)}]{Har15}
\bibinfo{author}{\bibfnamefont{J.~C.} \bibnamefont{Hardy}} \bibnamefont{and}
  \bibinfo{author}{\bibfnamefont{I.~S.} \bibnamefont{Towner}},
  \bibinfo{journal}{Phys. Rev. C} \textbf{\bibinfo{volume}{91}},
  \bibinfo{pages}{025501} (\bibinfo{year}{2015}).

\bibitem[{\citenamefont{Po\v{c}ani\'c et~al.}(2009)\citenamefont{Po\v{c}ani\'c,
  Alarcon, Alonzi, Baessler, Balascuta, Bowman, Bychkov, Byrne, Calarco,
  Cianciolo et~al.}}]{Poc09}
\bibinfo{author}{\bibfnamefont{D.}~\bibnamefont{Po\v{c}ani\'c}},
  \bibinfo{author}{\bibfnamefont{R.}~\bibnamefont{Alarcon}},
  \bibinfo{author}{\bibfnamefont{L.}~\bibnamefont{Alonzi}},
  \bibinfo{author}{\bibfnamefont{S.}~\bibnamefont{Baessler}},
  \bibinfo{author}{\bibfnamefont{S.}~\bibnamefont{Balascuta}},
  \bibinfo{author}{\bibfnamefont{J.}~\bibnamefont{Bowman}},
  \bibinfo{author}{\bibfnamefont{M.}~\bibnamefont{Bychkov}},
  \bibinfo{author}{\bibfnamefont{J.}~\bibnamefont{Byrne}},
  \bibinfo{author}{\bibfnamefont{J.}~\bibnamefont{Calarco}},
  \bibinfo{author}{\bibfnamefont{V.}~\bibnamefont{Cianciolo}},
  \bibnamefont{et~al.}, \bibinfo{journal}{Nucl. Instr. Meth. Phys. Res. A}
  \textbf{\bibinfo{volume}{611}}, \bibinfo{pages}{211 } (\bibinfo{year}{2009}).

\bibitem[{\citenamefont{Huyan et~al.}(2016)\citenamefont{Huyan,
  Naviliat-Cuncic, Bazin, Gade, Hughes, Liddick, Minamisono, Noji, Paulauskas,
  Simon et~al.}}]{Huy16}
\bibinfo{author}{\bibfnamefont{X.}~\bibnamefont{Huyan}},
  \bibinfo{author}{\bibfnamefont{O.}~\bibnamefont{Naviliat-Cuncic}},
  \bibinfo{author}{\bibfnamefont{D.}~\bibnamefont{Bazin}},
  \bibinfo{author}{\bibfnamefont{A.}~\bibnamefont{Gade}},
  \bibinfo{author}{\bibfnamefont{M.}~\bibnamefont{Hughes}},
  \bibinfo{author}{\bibfnamefont{S.}~\bibnamefont{Liddick}},
  \bibinfo{author}{\bibfnamefont{K.}~\bibnamefont{Minamisono}},
  \bibinfo{author}{\bibfnamefont{S.}~\bibnamefont{Noji}},
  \bibinfo{author}{\bibfnamefont{S.~V.} \bibnamefont{Paulauskas}},
  \bibinfo{author}{\bibfnamefont{A.}~\bibnamefont{Simon}},
  \bibnamefont{et~al.}, \bibinfo{journal}{Hyperfine Interact.}
  \textbf{\bibinfo{volume}{237}}, \bibinfo{pages}{1} (\bibinfo{year}{2016}).

\bibitem[{\citenamefont{Kofoed-Hansen}(1948)}]{Kof48}
\bibinfo{author}{\bibfnamefont{O.}~\bibnamefont{Kofoed-Hansen}},
  \bibinfo{journal}{Phys. Rev.} \textbf{\bibinfo{volume}{74}},
  \bibinfo{pages}{1785} (\bibinfo{year}{1948}).

\bibitem[{\citenamefont{Vetter et~al.}(2008)\citenamefont{Vetter, Abo-Shaeer,
  Freedman, and Maruyama}}]{Vet08}
\bibinfo{author}{\bibfnamefont{P.~A.} \bibnamefont{Vetter}},
  \bibinfo{author}{\bibfnamefont{J.~R.} \bibnamefont{Abo-Shaeer}},
  \bibinfo{author}{\bibfnamefont{S.~J.} \bibnamefont{Freedman}},
  \bibnamefont{and} \bibinfo{author}{\bibfnamefont{R.}~\bibnamefont{Maruyama}},
  \bibinfo{journal}{Phys. Rev. C} \textbf{\bibinfo{volume}{77}},
  \bibinfo{pages}{035502} (\bibinfo{year}{2008}), \eprint{0805.1212}.

\bibitem[{\citenamefont{Mund et~al.}(2013)\citenamefont{Mund, Maerkisch,
  Deissenroth, Krempel, Schumann, Abele, Petoukhov, and Soldner}}]{Mun13}
\bibinfo{author}{\bibfnamefont{D.}~\bibnamefont{Mund}},
  \bibinfo{author}{\bibfnamefont{B.}~\bibnamefont{Maerkisch}},
  \bibinfo{author}{\bibfnamefont{M.}~\bibnamefont{Deissenroth}},
  \bibinfo{author}{\bibfnamefont{J.}~\bibnamefont{Krempel}},
  \bibinfo{author}{\bibfnamefont{M.}~\bibnamefont{Schumann}},
  \bibinfo{author}{\bibfnamefont{H.}~\bibnamefont{Abele}},
  \bibinfo{author}{\bibfnamefont{A.}~\bibnamefont{Petoukhov}},
  \bibnamefont{and} \bibinfo{author}{\bibfnamefont{T.}~\bibnamefont{Soldner}},
  \bibinfo{journal}{Phys. Rev. Lett.} \textbf{\bibinfo{volume}{110}},
  \bibinfo{pages}{172502} (\bibinfo{year}{2013}).

\bibitem[{\citenamefont{Mendenhall et~al.}(2013)}]{Men13}
\bibinfo{author}{\bibfnamefont{M.~P.} \bibnamefont{Mendenhall}}
  \bibnamefont{et~al.} (\bibinfo{collaboration}{UCNA}), \bibinfo{journal}{Phys.
  Rev. C.} \textbf{\bibinfo{volume}{87}}, \bibinfo{pages}{032501}
  (\bibinfo{year}{2013}).

\bibitem[{\citenamefont{Johnson et~al.}(1963)\citenamefont{Johnson, Pleasonton,
  and Carlson}}]{Joh63}
\bibinfo{author}{\bibfnamefont{C.~H.} \bibnamefont{Johnson}},
  \bibinfo{author}{\bibfnamefont{F.}~\bibnamefont{Pleasonton}},
  \bibnamefont{and} \bibinfo{author}{\bibfnamefont{T.~A.}
  \bibnamefont{Carlson}}, \bibinfo{journal}{Phys. Rev.}
  \textbf{\bibinfo{volume}{132}}, \bibinfo{pages}{1149} (\bibinfo{year}{1963}).

\bibitem[{\citenamefont{Paul}(1970)}]{Pau70}
\bibinfo{author}{\bibfnamefont{H.}~\bibnamefont{Paul}}, \bibinfo{journal}{Nucl.
  Phys. A} \textbf{\bibinfo{volume}{154}}, \bibinfo{pages}{160 }
  (\bibinfo{year}{1970}).

\bibitem[{\citenamefont{Jackson et~al.}(1957)\citenamefont{Jackson, Treiman,
  and Wyld}}]{Jac57a}
\bibinfo{author}{\bibfnamefont{J.~D.} \bibnamefont{Jackson}},
  \bibinfo{author}{\bibfnamefont{S.~B.} \bibnamefont{Treiman}},
  \bibnamefont{and} \bibinfo{author}{\bibfnamefont{H.~W.} \bibnamefont{Wyld}},
  \bibinfo{journal}{Phys. Rev.} \textbf{\bibinfo{volume}{106}},
  \bibinfo{pages}{517} (\bibinfo{year}{1957}).

\bibitem[{\citenamefont{Konrad et~al.}(2012)\citenamefont{Konrad, Heil,
  Baessler, Po\v{c}ani{\'c}, and Gl{\"u}ck}}]{Kon10}
\bibinfo{author}{\bibfnamefont{G.}~\bibnamefont{Konrad}},
  \bibinfo{author}{\bibfnamefont{W.}~\bibnamefont{Heil}},
  \bibinfo{author}{\bibfnamefont{S.}~\bibnamefont{Baessler}},
  \bibinfo{author}{\bibfnamefont{D.}~\bibnamefont{Po\v{c}ani{\'c}}},
  \bibnamefont{and}
  \bibinfo{author}{\bibfnamefont{F.}~\bibnamefont{Gl{\"u}ck}},
  \emph{\bibinfo{title}{Impact of neutron decay experiments on non-standard
  model physics}} (\bibinfo{publisher}{World Scientific},
  \bibinfo{year}{2012}), pp. \bibinfo{pages}{660--672}.

\bibitem[{\citenamefont{Wauters et~al.}(2014)\citenamefont{Wauters,
  Garc\'{\i}a, and Hong}}]{Wau14}
\bibinfo{author}{\bibfnamefont{F.}~\bibnamefont{Wauters}},
  \bibinfo{author}{\bibfnamefont{A.}~\bibnamefont{Garc\'{\i}a}},
  \bibnamefont{and} \bibinfo{author}{\bibfnamefont{R.}~\bibnamefont{Hong}},
  \bibinfo{journal}{Phys. Rev. C} \textbf{\bibinfo{volume}{89}},
  \bibinfo{pages}{025501} (\bibinfo{year}{2014}), \bibinfo{note}{[Erratum:
  Phys. Rev. C {\bf 91}, 049904 (2015)]}.

\bibitem[{\citenamefont{Chirovsky et~al.}(1984)\citenamefont{Chirovsky, Lee,
  Sabbas, Becker, Groves, and Wu}}]{Chi84}
\bibinfo{author}{\bibfnamefont{L.~M.} \bibnamefont{Chirovsky}},
  \bibinfo{author}{\bibfnamefont{W.-P.} \bibnamefont{Lee}},
  \bibinfo{author}{\bibfnamefont{A.~M.} \bibnamefont{Sabbas}},
  \bibinfo{author}{\bibfnamefont{A.~J.} \bibnamefont{Becker}},
  \bibinfo{author}{\bibfnamefont{J.~L.} \bibnamefont{Groves}},
  \bibnamefont{and} \bibinfo{author}{\bibfnamefont{C.}~\bibnamefont{Wu}},
  \bibinfo{journal}{Nucl. Instr. Meth. Phys. Res.}
  \textbf{\bibinfo{volume}{219}}, \bibinfo{pages}{103 } (\bibinfo{year}{1984}).

\bibitem[{\citenamefont{Wietfeldt et~al.}(2009)\citenamefont{Wietfeldt, Byrne,
  Collett, Dewey, Jones, Komives, Laptev, Nico, Noid, Stephenson
  et~al.}}]{Wie09}
\bibinfo{author}{\bibfnamefont{F.}~\bibnamefont{Wietfeldt}},
  \bibinfo{author}{\bibfnamefont{J.}~\bibnamefont{Byrne}},
  \bibinfo{author}{\bibfnamefont{B.}~\bibnamefont{Collett}},
  \bibinfo{author}{\bibfnamefont{M.}~\bibnamefont{Dewey}},
  \bibinfo{author}{\bibfnamefont{G.}~\bibnamefont{Jones}},
  \bibinfo{author}{\bibfnamefont{A.}~\bibnamefont{Komives}},
  \bibinfo{author}{\bibfnamefont{A.}~\bibnamefont{Laptev}},
  \bibinfo{author}{\bibfnamefont{J.}~\bibnamefont{Nico}},
  \bibinfo{author}{\bibfnamefont{G.}~\bibnamefont{Noid}},
  \bibinfo{author}{\bibfnamefont{E.}~\bibnamefont{Stephenson}},
  \bibnamefont{et~al.}, \bibinfo{journal}{Nucl. Instr. Meth. Phys. Res. A}
  \textbf{\bibinfo{volume}{611}}, \bibinfo{pages}{207 } (\bibinfo{year}{2009}).

\bibitem[{\citenamefont{Grigoriev et~al.}(1968)\citenamefont{Grigoriev,
  Grishin, Vladimirsky, and Nikolaevsky}}]{Gri68}
\bibinfo{author}{\bibfnamefont{V.~K.} \bibnamefont{Grigoriev}},
  \bibinfo{author}{\bibfnamefont{A.~P.} \bibnamefont{Grishin}},
  \bibinfo{author}{\bibfnamefont{V.~V.} \bibnamefont{Vladimirsky}},
  \bibnamefont{and} \bibinfo{author}{\bibfnamefont{E.~S.}
  \bibnamefont{Nikolaevsky}}, \bibinfo{journal}{Sov. J. Nucl. Phys.}
  \textbf{\bibinfo{volume}{6}}, \bibinfo{pages}{239} (\bibinfo{year}{1968}).

\bibitem[{\citenamefont{Gl{\"u}ck}(1998)}]{Glu98}
\bibinfo{author}{\bibfnamefont{F.}~\bibnamefont{Gl{\"u}ck}},
  \bibinfo{journal}{Nucl. Phys. A} \textbf{\bibinfo{volume}{628}},
  \bibinfo{pages}{493 } (\bibinfo{year}{1998}).

\bibitem[{\citenamefont{Severijns and Naviliat-Cuncic}(2011)}]{Sev11}
\bibinfo{author}{\bibfnamefont{N.}~\bibnamefont{Severijns}} \bibnamefont{and}
  \bibinfo{author}{\bibfnamefont{O.}~\bibnamefont{Naviliat-Cuncic}},
  \bibinfo{journal}{Annu. Rev. Nucl. Part. Sci.} \textbf{\bibinfo{volume}{61}},
  \bibinfo{pages}{23} (\bibinfo{year}{2011}).

\bibitem[{\citenamefont{Pitcairn et~al.}(2009)}]{Pit08}
\bibinfo{author}{\bibfnamefont{J.~R.~A.} \bibnamefont{Pitcairn}}
  \bibnamefont{et~al.}, \bibinfo{journal}{Phys. Rev. C}
  \textbf{\bibinfo{volume}{79}}, \bibinfo{pages}{015501}
  (\bibinfo{year}{2009}), \eprint{0811.0052}.

\bibitem[{\citenamefont{Adelberger et~al.}(1999)\citenamefont{Adelberger,
  Ortiz, Garc\'{\i}a, Swanson, Beck, Tengblad, Borge, Martel, Bichsel, and
  ISOLDE~Collaboration}}]{Ade99}
\bibinfo{author}{\bibfnamefont{E.~G.} \bibnamefont{Adelberger}},
  \bibinfo{author}{\bibfnamefont{C.}~\bibnamefont{Ortiz}},
  \bibinfo{author}{\bibfnamefont{A.}~\bibnamefont{Garc\'{\i}a}},
  \bibinfo{author}{\bibfnamefont{H.~E.} \bibnamefont{Swanson}},
  \bibinfo{author}{\bibfnamefont{M.}~\bibnamefont{Beck}},
  \bibinfo{author}{\bibfnamefont{O.}~\bibnamefont{Tengblad}},
  \bibinfo{author}{\bibfnamefont{M.~J.~G.} \bibnamefont{Borge}},
  \bibinfo{author}{\bibfnamefont{I.}~\bibnamefont{Martel}},
  \bibinfo{author}{\bibfnamefont{H.}~\bibnamefont{Bichsel}}, \bibnamefont{and}
  \bibinfo{author}{\bibfnamefont{t.}~\bibnamefont{ISOLDE~Collaboration}},
  \bibinfo{journal}{Phys. Rev. Lett.} \textbf{\bibinfo{volume}{83}},
  \bibinfo{pages}{3101} (\bibinfo{year}{1999}).

\bibitem[{\citenamefont{Gorelov et~al.}(2005)\citenamefont{Gorelov, Melconian,
  Alford, Ashery, Ball, Behr, Bricault, D'Auria, Deutsch, Dilling
  et~al.}}]{Gor05}
\bibinfo{author}{\bibfnamefont{A.}~\bibnamefont{Gorelov}},
  \bibinfo{author}{\bibfnamefont{D.}~\bibnamefont{Melconian}},
  \bibinfo{author}{\bibfnamefont{W.~P.} \bibnamefont{Alford}},
  \bibinfo{author}{\bibfnamefont{D.}~\bibnamefont{Ashery}},
  \bibinfo{author}{\bibfnamefont{G.}~\bibnamefont{Ball}},
  \bibinfo{author}{\bibfnamefont{J.~A.} \bibnamefont{Behr}},
  \bibinfo{author}{\bibfnamefont{P.~G.} \bibnamefont{Bricault}},
  \bibinfo{author}{\bibfnamefont{J.~M.} \bibnamefont{D'Auria}},
  \bibinfo{author}{\bibfnamefont{J.}~\bibnamefont{Deutsch}},
  \bibinfo{author}{\bibfnamefont{J.}~\bibnamefont{Dilling}},
  \bibnamefont{et~al.}, \bibinfo{journal}{Phys. Rev. Lett.}
  \textbf{\bibinfo{volume}{94}}, \bibinfo{pages}{142501}
  (\bibinfo{year}{2005}).

\bibitem[{\citenamefont{Gorelov}(2008)}]{Gor08}
\bibinfo{author}{\bibfnamefont{A.~I.} \bibnamefont{Gorelov}}
  (\bibinfo{year}{2008}), \bibinfo{note}{phD Thesis, Simon Fraser University}.

\bibitem[{\citenamefont{Garc\'{i}a}(2016)}]{Gar16}
\bibinfo{author}{\bibfnamefont{A.}~\bibnamefont{Garc\'{i}a}}
  (\bibinfo{year}{2016}), \bibinfo{note}{private communication}.

\bibitem[{\citenamefont{Gorelov}(2016)}]{Gor16}
\bibinfo{author}{\bibfnamefont{A.~I.} \bibnamefont{Gorelov}}
  (\bibinfo{year}{2016}), \bibinfo{note}{private communication}.

\bibitem[{\citenamefont{Mougeot and Bisch}(2014)}]{Mou14}
\bibinfo{author}{\bibfnamefont{X.}~\bibnamefont{Mougeot}} \bibnamefont{and}
  \bibinfo{author}{\bibfnamefont{C.}~\bibnamefont{Bisch}},
  \bibinfo{journal}{Phys. Rev. A} \textbf{\bibinfo{volume}{90}},
  \bibinfo{pages}{012501} (\bibinfo{year}{2014}).

\bibitem[{\citenamefont{Golovko et~al.}(2008)\citenamefont{Golovko, Iacob, and
  Hardy}}]{Gol08}
\bibinfo{author}{\bibfnamefont{V.}~\bibnamefont{Golovko}},
  \bibinfo{author}{\bibfnamefont{V.}~\bibnamefont{Iacob}}, \bibnamefont{and}
  \bibinfo{author}{\bibfnamefont{J.}~\bibnamefont{Hardy}},
  \bibinfo{journal}{Nucl. Instr. Meth. Phys. Res. A}
  \textbf{\bibinfo{volume}{594}}, \bibinfo{pages}{266 } (\bibinfo{year}{2008}).

\bibitem[{\citenamefont{Soti et~al.}(2013)\citenamefont{Soti, Wauters,
  Breitenfeldt, Finlay, Kraev, Knecht, Porobi\'{c}, Z\'{a}kouck\'{y}, and
  Severijns}}]{Sot13}
\bibinfo{author}{\bibfnamefont{G.}~\bibnamefont{Soti}},
  \bibinfo{author}{\bibfnamefont{F.}~\bibnamefont{Wauters}},
  \bibinfo{author}{\bibfnamefont{M.}~\bibnamefont{Breitenfeldt}},
  \bibinfo{author}{\bibfnamefont{P.}~\bibnamefont{Finlay}},
  \bibinfo{author}{\bibfnamefont{I.}~\bibnamefont{Kraev}},
  \bibinfo{author}{\bibfnamefont{A.}~\bibnamefont{Knecht}},
  \bibinfo{author}{\bibfnamefont{T.}~\bibnamefont{Porobi\'{c}}},
  \bibinfo{author}{\bibfnamefont{D.}~\bibnamefont{Z\'{a}kouck\'{y}}},
  \bibnamefont{and}
  \bibinfo{author}{\bibfnamefont{N.}~\bibnamefont{Severijns}},
  \bibinfo{journal}{Nucl. Instr. Meth. Phys. Res. A}
  \textbf{\bibinfo{volume}{728}}, \bibinfo{pages}{11 } (\bibinfo{year}{2013}).

\bibitem[{\citenamefont{Knecht et~al.}(2013)\citenamefont{Knecht, Alexander,
  Bagdasarova, Cope, Delbridge, Fl\'echard, Garc\'ia, Hong, Li\'enard, Mueller
  et~al.}}]{Kne13}
\bibinfo{author}{\bibfnamefont{A.}~\bibnamefont{Knecht}},
  \bibinfo{author}{\bibfnamefont{Z.~T.} \bibnamefont{Alexander}},
  \bibinfo{author}{\bibfnamefont{Y.}~\bibnamefont{Bagdasarova}},
  \bibinfo{author}{\bibfnamefont{T.~M.} \bibnamefont{Cope}},
  \bibinfo{author}{\bibfnamefont{B.~G.} \bibnamefont{Delbridge}},
  \bibinfo{author}{\bibfnamefont{X.}~\bibnamefont{Fl\'echard}},
  \bibinfo{author}{\bibfnamefont{A.}~\bibnamefont{Garc\'ia}},
  \bibinfo{author}{\bibfnamefont{R.}~\bibnamefont{Hong}},
  \bibinfo{author}{\bibfnamefont{E.}~\bibnamefont{Li\'enard}},
  \bibinfo{author}{\bibfnamefont{P.}~\bibnamefont{Mueller}},
  \bibnamefont{et~al.}, \bibinfo{journal}{AIP Conf. Proc.}
  \textbf{\bibinfo{volume}{1560}}, \bibinfo{pages}{636} (\bibinfo{year}{2013}).

\end{thebibliography}
%%%%%%%%%%%%%%%%%%%%%%%%%%%%%%%%%%%%%%%%%%%%%%%%%%%%%%%%%%%%%%%%%%%%%%%%%%%%%%%%
\end{document}